\documentclass[journal]{IEEEtran}

\usepackage{algorithm}
\usepackage{algorithmicx}
\usepackage[noend]{algpseudocode} 
\usepackage[hidelinks]{hyperref}
\hypersetup{
  colorlinks   = true, 
  urlcolor     = blue, 
  linkcolor    = blue, 
  citecolor   = green 
}
\usepackage{cite}
\usepackage{acro}
\usepackage{graphicx}
\usepackage{float}
\usepackage{xcolor}
\usepackage{amsmath}
\usepackage{array}
\usepackage[utf8x]{inputenc} 
\usepackage{pifont}
\newcommand{\checkmark}{\ding{51}}%
\newcommand{\xmark}{\ding{55}}%

\begin{document}

\title{Holistic Fine-grained GGS Characterization: From Detection to Unbalanced Classification}
\author{Yuzhe Lu, Haichun~Yang, Zuhayr Asad, Zheyu Zhu, Tianyuan Yao, Jiachen Xu, Agnes B. Fogo, and Yuankai Huo, \IEEEmembership{Senior Member, IEEE}

\thanks{This work was supported by NIH NIDDK DK56942(ABF). \emph{(Corresponding author: Yuankai Huo. Email: yuankai.huo@vanderbilt.edu)}}

\thanks{Y. Lu, Z. Asad, Z. Zhu, T. Yao, J. Xu, Y. Huo were with the Department of Electrical Engineering and Computer Science, Vanderbilt University, Nashville, TN 37235 USA.}

\thanks{H. Yang, A. B. Fogo were with the Department
of Pathology, Vanderbilt University Medical Center, Nashville,
TN, 37215, USA.}
}

\maketitle

\markboth{Manuscript pre-print, January~2022}%
{Shell \MakeLowercase{\textit{et al.}}: Bare Demo of IEEEtran.cls for IEEE Journals}

\maketitle

\section{Abstract}
\noindent\textbf{Purpose.} Recent studies have demonstrated the diagnostic and prognostic values of global glomerulosclerosis (GGS) in IgA nephropathy, aging, and end-stage renal disease. However, the fine-grained quantitative analysis of multiple GGS subtypes (e.g., obsolescent, solidified, and disappearing glomerulosclerosis) is typically a resource extensive manual process. Very few automatic methods, if any, have been developed to bridge this gap for such analytics. In this paper, we present a holistic pipeline to quantify GGS (with both detection and classification) from a whole slide image in a fully automatic manner. In addition, we conduct the fine-grained classification for the sub-types of GGS. Our study releases the open-source quantitative analytical tool for fine-grained GGS characterization while tackling the technical challenges in unbalanced classification and integrating detection and classification. \\\textbf{Approach.} We present a deep learning-based framework to perform fine-grained detection and classification of global glomerulosclerosis, with a hierarchical two-stage design. Moreover, we incorporate the state-of-the-art transfer learning techniques to achieve a more generalizable deep learning model for tackling the imbalanced distribution of our dataset. This way, we build a highly efficient WSI-to-results global glomerulosclerosis characterization pipeline. Meanwhile, we investigated the largest fine-grained GGS cohort as of yet with 11,462 glomeruli and 10,619 non-glomeruli, which include 7,841 globally sclerotic glomeruli of three distinct categories. With this data, we apply deep learning techniques to achieve 1) fine-grained GGS characterization 2) GGS vs non-GGS classification 3) improved glomeruli detection results. \\\textbf{Results.} For fine-grained global glomerulosclerosis characterization, when pre-trained on the larger dataset, our model can achieve a $0.778$ macro $F1$ score, compared to a $0.746$ macro $F1$ score when using the regular ImageNet-pretrained weights. On the external dataset, our best model achieves an AUC score of $0.994$ when tasked with differentiating GGS from normal glomeruli. Using our dataset, we are able to build algorithms that allow for fine-grained classification of glomeruli lesions and are robust to distribution shifts.\\\textbf{Conclusion.} Our study showed that the proposed methods consistently improve the detection and fine-grained classification performance through both cross-validation and external validation. Our code and pretrained models have been released for public use at \url{https://github.com/luyuzhe111/glomeruli}.

\begin{figure*} [t]
  \begin{center}

  \includegraphics[width=0.8\textwidth]{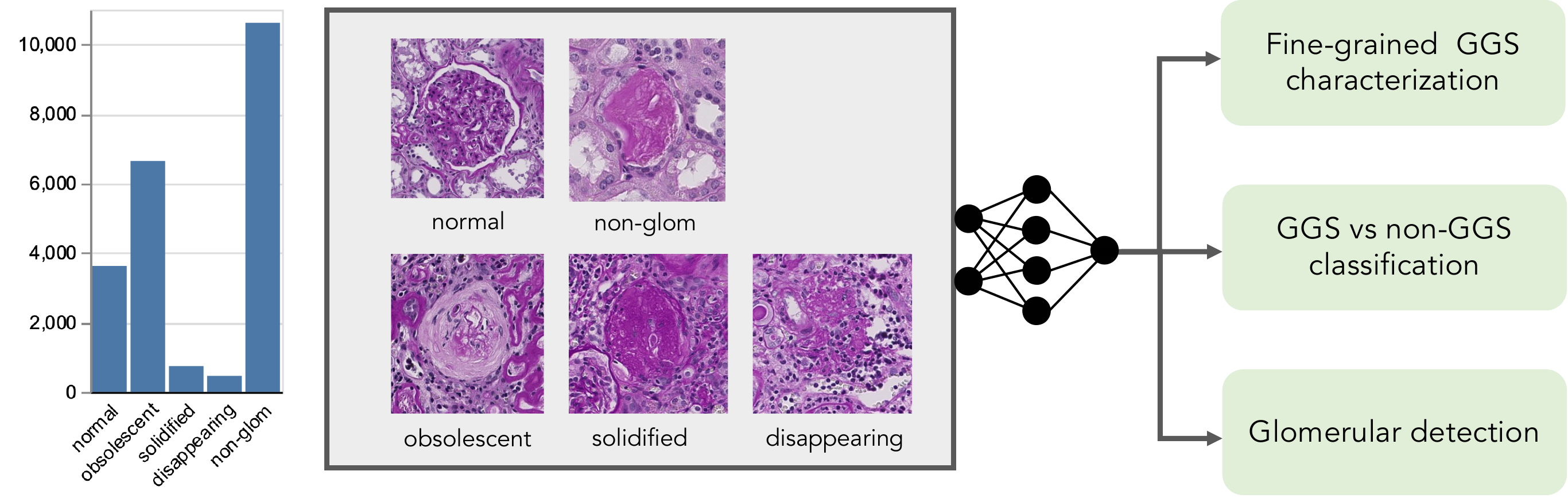}
 
  \end{center}
  \caption  
  {  
     This figure shows examples of fine-grained glomerulus definitions as well as their highly unbalanced distribution, where there are a large amount of non-glomerulus patches, plenty of obsolescent and normal patches, but fewer solidified and disappearing patches.
  }
\end{figure*}

\section{Introduction}
\IEEEPARstart{T}{he} identification of non-sclerotic and sclerotic glomeruli is an essential task in clinical renal pathology and scientific kidney research as a quantitative measurement corresponding to various critical clinical outcomes~\cite{benchimol2003focal}. Global glomerulosclerosis (GGS, scarring lesion or hyaline deposition $>$ 50$\%$ in a glomerulus) is historically less investigated as compared with segmental glomerulosclerosis (scarring lesion or hyaline deposition $<$ 50 $\%$ in a glomerulus) in clinical guidelines, since the earlier stages of glomerulosclerosis (e.g., Focal  Segmental glomerulosclerosis) has been widely accepted as an essential clinical marker for various chronic kidney diseases. In the past few years, more and more studies have demonstrated the diagnostic and prognostic values of GGS in IgA nephropathy~\cite{tan2020global}, aging~\cite{chung2020age}, and end-stage renal disease~\cite{hommos2018global}. 
The quantitative analysis of GGS (e.g., counting the percentage of glomeruli with GGS) is typically a resource extensive manual process, including glomerular detection (localizing all glomeruli from a whole slide image (WSI)) and classification (classifying each glomerulus to GGS subtypes). However, a holistic computational solution to tackle the resource-intensive fine-grained GGS characterization does not yet exist.

In more fine-grained definitions~\cite{tan2020global}, GGS can be further classified into three categories: obsolescent glomerulosclerosis, solidified glomerulosclerosis, and disappearing glomerulosclerosis\cite{marcantoni2002hypertensive}. The fine-grained glomerulosclerosis phenotype could provide more precise evidence to support both scientific research and clinical decision-making. For example, solidified glomerulosclerosis is identified as a key prognostic indicator in end-stage renal disease (ESRD)\cite{zhao2021solidified}. Higher frequencies of solidified and disappearing glomerulosclerosis and lower frequencies of obsolescent glomerulosclerosis are shown to be correlated with apolipoprotein L1 gene (APOL1) related chronic kidney disease (CKD)\cite{larsen2015histopathologic}. However, differentiating these patterns typically requires heavy manual efforts by trained clinical experts, which is not only tedious, but also labor-intensive. Therefore, there is a strong need to develop automatic and holistic detection and classification algorithms to perform fine-grained glomerulosclerosis classification, especially with an increasingly large amount of digitized data from WSIs.

In this study, we propose a holistic and fully automatic deep learning pipeline to achieve dense glomerular detection and classification on fine-grained GGS subtypes. To do so, we collect a large-scale glomeruli dataset labeled by an experienced renal pathologist. The dataset consists of 10,840 glomeruli from one of the four classes: normal, global obsolescent, global solidified, and global disappearing. Along the way, the pathologist also labeled 10,619 image patches that did not contain a glomerulus, but rather tissues that were considered glomeruli (false positive) from our detection algorithm~\cite{yang2020circlenet}. We employ a transfer learning framework to improve the classification of GGS subtypes using the recently released models that are pretrained on datasets of a much larger scale (orders of magnitude larger than ImageNet~\cite{deng2009imagenet}). Together with other technical innovations for unbalanced classification, our model achieves a $0.778$ macro $F1$ score to perform fine-grained GGS characterization. On the external dataset, our best model achieves a perfect AUC score of $0.994$ to identify GGS from normal glomeruli.

Our contribution is in four fold:

\begin{itemize}
    \item We propose a holistic and fully automatic deep learning pipeline to achieve dense glomerular detection and classification with fine-grained GGS subtypes.
    \item The proposed method integrates transfer learning, hard negative mining, and unbalanced training to overcome the technical hurdles in learning from an unbalanced GSS dataset.
    \item To the best of our knowledge, this is by far the largest study (10,840 glomeruli) that investigates fine-grained GSS classification with three subtypes: obsolescent glomerulosclerosis, solidified glomerulosclerosis, and disappearing glomerulosclerosis.
	\item We present a comprehensive evaluation on both internal and external validation (one public external dataset and two in-house datasets) spanning binary classification, multi-class classification, and object detection.
	
\end{itemize}

\section{Related Works}
\subsection{Glomerulosclerosis Classification}
The success of deep convolutional neural networks (CNNs) in recent years has spurred extensive research on their application in renal pathology, where detection and classification of glomeruli are critical for quantitative evaluation and precise diagnosis. Previously, Gallego et al.~\cite{gallego2018glomerulus} used CNN-based classification, while Gloria et al.~\cite{bueno2020glomerulosclerosis} utilized semantic segmentation, to achieve glomeruli detection in whole slide images. More recently, Yang et al. proposed an anchor-free detection strategy using circle representation~\cite{yang2020circlenet} that is optimized for round glomeruli and demonstrates superior performance. 

On the other front, many studies have been conducted to classify different glomerular lesions using computer-aided approaches~\cite{marsh2018deep,zeng2020identification,ginley2020fully}. Specifically, Uchino et al.~\cite{uchino2020classification} trained deep learning algorithms to classify several renal pathological findings and achieved an AUC score of 0.983 for global sclerosis characterization. However, most algorithms are tested using in-house data and thus, their performances can be highly volatile in the face of distribution shift. Moreover, there are few, if any, studies that have developed deep learning approaches to classify globally sclerotic glomeruli into three fine-grained categories: obsolescent, solidified, and disappearing glomerulosclerosis. Such fine-grained characterization is challenging as the available data is rare and their distribution is highly imbalanced. For instance, the presence of obsolescent glomerulosclerosis is naturally much higher than solidified or disappearing glomerulosclerosis, leading to the technical difficulty that is well known as the imbalanced classes problem. 

\begin{figure*}
  \begin{center}

  \includegraphics[width=0.9\textwidth]{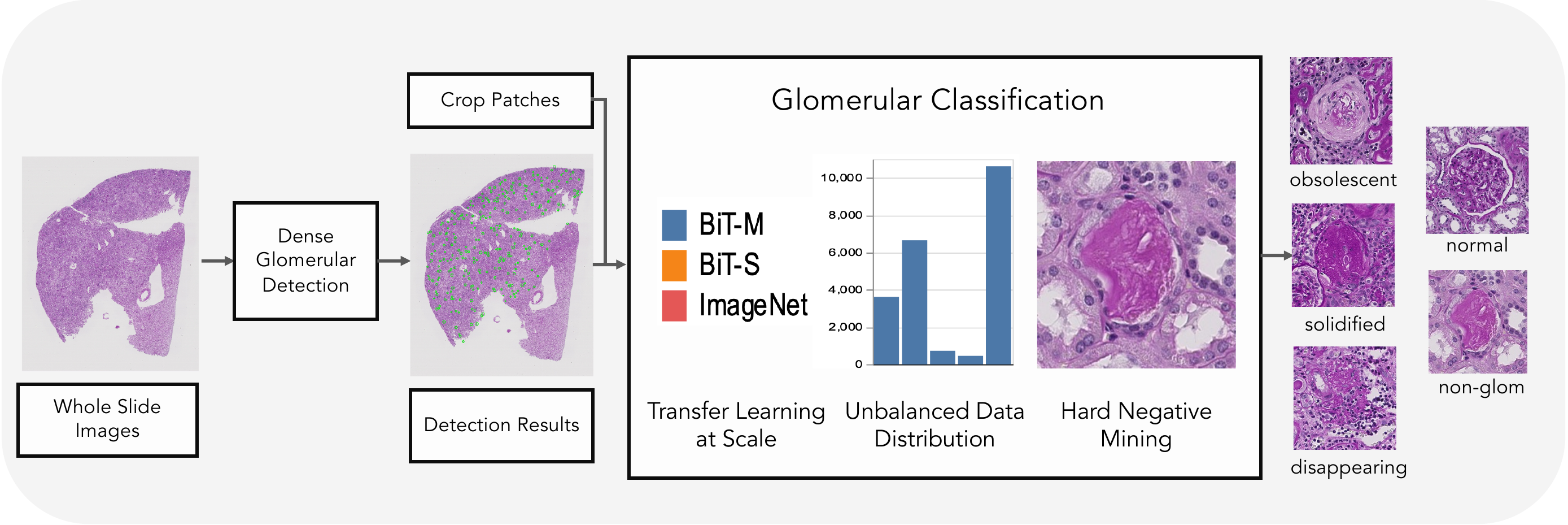}
 
  \end{center}
  \caption{  
  This figure shows the methodologies utilized in our proposed holistic fine-grained GGS characterization pipeline. We leverage transfer learning at various scales to improve GGS classification results on both in-house and external data. As our collected dataset is naturally unbalanced, several unbalanced learning strategies are investigated to address this issue. In addition, we perform hard negative mining to further improve glomeruli detection results using our trained classifier.
  }
\end{figure*} 

\subsection{Transfer Learning for Medical Imaging}

Transfer learning is a commonly used strategy in building deep learning algorithms for medical imaging applications, given that labeled medical data is often rare. The standard practice in the medical imaging community has been to initialize a classical model architecture with weights pretrained on ImageNet\cite{deng2009imagenet}. Though this practice has been recently challenged \cite{raghu2019transfusion}, later studies\cite{kolesnikov2020big} have demonstrated that models pretrained on larger datasets transfer better to downstream tasks. Notably, Mustafa et al.\cite{mustafa2021supervised} showed that there are consistent improvements for in-domain performance, generalization under domain-shift, and data efficiency in three downstream medical image classification tasks (mammography, chest X-Ray, and dermatology) when the models are pretrained on larger natural image datasets. It is expected that applying those pretrained weights would also benefit our algorithm to classify the glomeruli patches.

\section{Method}

The methodology followed in this study can be broken down into three major parts: 1) transfer learning for glomerulosclerosis, 2) hard negative mining, and 3) holistic detection and classification pipeline. In classification, we leverage transfer learning at scale to perform fine-grained characterization of globally sclerotic glomeruli. By utilizing hard negative cases in training, our model is able to effectively identify non-glomerulus patches. Equipped with these two capabilities, we present a holistic glomeruli detection-classification deep learning pipeline for renal pathology.

\subsection{Transfer Learning for Glomerulosclerosis}
In this section, we aim to perform fine-grained characterizations of global sclerosis by classifying it into three categories: obsolescent glomerulosclerosis, solidified glomerulosclerosis, and disappearing glomerulosclerosis. 

To deal with the challenge of limited labeled data, we leverage the family of Big Transfer Models (BiT)~\cite{kolesnikov2020big} that are pre-trained on natural image datasets of different scales and evaluate their performances. Compared with the default ResNet\cite{he2016identity} implementations in PyTorch, BiT models utilize the ResNet-V2 architecture of varying sizes and substitute Batch Norm with a combination of Group Norm\cite{wu2018group} and Weight Standardization\cite{qiao2019weight} to maximize the benefits of transfer learning. Recent work has shown that these models are highly successful across different computer vision and medical imaging tasks. Currently, models that are pre-trained on ILSVRC-2012\cite{russakovsky2015imagenet}-which contains 1.3M images (BiT-S), and ImageNet-21k\cite{deng2009imagenet}-which contains 14M images (BiT-M), are released for public use. 

Moreover, as the distribution of globally sclerotic glomeruli is highly imbalanced in nature, we adopt two measures to tackle this issue. Firstly, minority cases are sampled more frequently such that they have roughly equal representation to majority classes within each mini-batch. In addition, to weaken the contribution of easy cases to the loss function, we use the focal loss \cite{lin2017focal} with $\gamma=2.5$. 

\subsection{Hard Negative Mining}
In object detection, hard negative mining\cite{sung1996learning, felzenszwalb2010cascade, jin2018unsupervised} is often proposed to deal with the data imbalance problem, where detectors usually have to evaluate $10^4$ to $10^5$ candidate regions per image for only a few relevant locations containing the desired objects. One prominent issue is that most of these candidate regions are easy examples from which no useful signal can be learned. To successfully train the model, it is critical to utilize those so-called hard examples. In our use case, the hard negative mining was performed as the following steps. First, we applied the open-source CircleNet\cite{yang2020circlenet} model to WSIs to get patches that are believed to contain glomerulus. Then, an experienced renal pathologist inspected each of these patches and identified ones that don't contain a glomerulus. As this process is rather time-consuming, only a single pass was done. In the end, 10,619 non-glomerulus images were collected. We hypothesize that these false positives are conceptually equivalent to hard negatives in this case and mining these data can contribute to our classifier's ability to identity non-glomerulus images.

\subsection{Holistic Detection-Then-Classification Pipeline}

We propose a holistic glomeruli detection and classification pipeline by integrating our classifier as a post-processing step for glomeruli detection algorithms, where non-glomerulus patches will be identified and filtered out. Closely related to our proposed pipeline is work by Jha et al.\cite{jha2021instance}, where the authors show that for high-resolution medical imaging data, a detect-then-segment pipeline performs better than the de facto standard end-to-end segmentation pipeline such as the Mask-RCNN\cite{he2017mask}. Thus, it is promising to adapt such a ``two-stage" design, where the inputs of classification on high-resolution glomeruli images are cropped from high-resolution WSIs instead of low-resolution feature maps that are used for the detection. In detection algorithms, a required step is to find the optimal threshold that strikes a balance between finding all true positives and avoiding false positives. As excessive false positives can be a huge burden to clinical experts who are usually tasked with manual quality assurance, the threshold is often chosen such that some true positives are unavoidably left out. However, by adding a classifier that efficiently distinguishes glomerulus versus non-glomerulus for the glomeruli detection algorithm as a post-processing step, we show that we can effectively improve the detection performance by lowering the detection threshold without adding extra manual efforts.

\section{Experiments}

\subsection{Data}

Non-cancer regions of nephrectomy were acquired from 157 patients with kidney cancer (male108, female 49, age 60.2$\pm$14.8). Clinically, 42 of 157 were diagnosed with diabetes while 50 of 157 had hypertension. Pathology diagnosis included 9 cases of arteriolonephrosclerosis, 1 case of focal segmental glomerulosclerosis, 7 cases of diabetic nephropathy, 2 cases of acute tubular necrosis, and 7 cases of moderate to severe interstitial fibrosis. 

The tissues were routinely processed and paraffin-embedded, with 3 μm thickness sections cut and stained with PAS. The whole slide image was scanned (40X) for each case by using Leica SCN400 Slide Scanner, and glomeruli in the entire slide were analyzed. Due to the fact that WSIs often have high resolution and thus are typically too large to be directly used for training deep learning models, it is a common practice to crop the whole images into small patches for deep learning tasks\cite{bueno2020data, khened2021generalized}. 22,081 glomeruli were extracted from WSI using the EasierPath semi-manual annotation software\cite{zhu2020easierpath}. Then, all glomeruli were manually labeled by a renal pathologist, including 3,621 normal glomeruli, 6,647 global obsolescent glomeruli, 735 global solidified glomeruli, 459 global disappearing glomeruli, and 10,619 non-glomeruli. These glomeruli patches, which are cropped from the original WSIs (0.25 \textmu m per pixel), are resized via downsampling from 512$\times$512-pixel to 256$\times$256-pixel\cite{bao2020bifocal} for training, validation, and testing. The data were de-identified, and studies were approved by the Institutional Review Board (IRB).

An external glomeruli dataset\cite{bueno2020data} was also used to evaluate the performance of our model. The dataset consists of 2,340 PNG images containing a single glomerulus. Among them, 1,170 glomeruli are normal and 1,170 glomeruli are sclerosed. As this dataset was collected from a completely different population, it can be used to demonstrate the robustness of our algorithm to distribution shifts. 

\subsection{Models}
\subsubsection{Baseline}
For the fine-grained classification experiments, the classical ResNet architecture with batch normalization was used as the baseline. The ImageNet\cite{deng2009imagenet} pretrained classification networks have been broadly used for downstream medical imaging tasks. We used both ResNet50 and ResNet101 pretrained on ILSVRC2012 ImageNet as our baseline models.

\subsubsection{BiT Models}
The family of BiT (Big Transfer) models were based on variants of ResNet-V2 architectures. Compared to the baseline architecture, ResNet-V2 substituted batch normalized for group normalization and weight standardization as pretrained models. The BiT-S model was trained on the regular ILSVRC2012 ImageNet dataset (consisting of 1.3 million natural images), the BiT-M model was trained on ImageNet-21k (consisting of 14 million natural images), and the BiT-L model was trained on JFT-300M (consisting of 300M images). Currently, only the BiT-S and BiT-M models were released for public use. Thus, the BiT models were used in this study to evaluate the impact of transfer learning, including ResNet50-BiT-S, ResNet50-BiT-M, ResNet101-BiT-S, and ResNet101-BiT-M.

\subsection{Experiment Design}
We adapted the models mentioned above by customizing the fully connected layers based on our tasks while retaining all other pretrained weights. The model was trained and tested with standard five-fold cross-validation. Briefly, the data was split into five folds at the patient level, where each fold was withheld as testing data once. The remaining data for each fold was split as 75\% training data and 25\% validation data. Therefore, for each fold, the final split was 60\% training, 20\% validation, and 20\% test. To avoid data contamination, all glomeruli from each patient were used either for training, validation, or testing. 

The model was trained using focal loss and the Adam optimizer. For ResNet50 models, we used a batch size of 16 and a learning rate of 1e-4. For ResNet101 models, we also used a batch size of 16 but a smaller learning rate of 1e-5. The $\gamma$ parameter for focal loss was set to 2.5 for all experiments.

\subsection{Evaluation Metrics and Statistical Methods}

As our dataset was highly imbalanced, the balanced $F1$ score \cite{chinchor1993muc} was used to evaluate the performance. F1 score is defined as the harmonic mean of precision and recall, and the balanced $F1$ was calculated as the arithmetic mean of each class's $F1$ score (one vs. remaining). In each fold, the model with the best balanced accuracy in the validation set was selected for testing. To generate statistically stable results, we ran 5-fold cross-validation five times with different seeds. The mean and standard deviation were shown on data tables and plots when applicable. The mean was reported instead of median as no outlier was identified.

\section{Results}
Our results presented in this section are divided into three main components in order to explore each aspect of our experimentation: 1) fine-grained classification of GGS, 2) external validation of GGS classification, and 3) integration with glomeruli detection algorithms.

\subsection{Cross-validation}
The cross-validation results in Table 1 demonstrated two main points. Firstly, replacing batch normalization for a combination of weight standardization and group normalization helped the transfer learning performance, as both the BiT-S model and the default PyTorch model used weights pretrained on ILSVRC-2012. The only difference between these two kinds of models was that the default PyTorch model used batch normalization while the BiT models used weight standardization and group normalization. Secondly, transfer learning at a larger scale consistently brought better performance measured by balanced $F1$ score. Overall, the best balanced $F1$ score was given by using ResNet101-BiT-M. 

\begin{table*}
\caption{Cross validation results for the internal data measured by $F1$ score.}
\medskip
\resizebox{\textwidth}{!}{%
\begin{tabular}{l c c c c c c}
\hline
& Normal & Obsolescent & Solidified & Disappearing & Non-glom & Macro $F1$\\ 
\hline
ResNet50-RandInit & 
0.932 ± 0.019 & 0.825 ± 0.002 & 0.438 ± 0.012 & 0.266 ± 0.024 & 0.892 ± 0.013 & 0.670 ± 0.011\\ 

ResNet50-ImageNet & 
0.970 ± 0.005 & 0.884 ± 0.005 & 0.541 ± 0.016 & 0.341 ± 0.022 & 0.953 ± 0.004 & 0.738 ± 0.005\\ 

ResNet50-BiT-S 
& 0.976 ± 0.005 & 0.902 ± 0.006 & 0.562 ± 0.015 & 0.388 ± 0.030 & 0.958 ± 0.006 & 0.757 ± 0.01\\ 

ResNet50-BiT-M 
& 0.976 ± 0.005 & 0.900 ± 0.013 & 0.565 ± 0.029 & 0.375 ± 0.041  & 0.960 ± 0.008 & 0.755 ± 0.012\\ 
\hline

ResNet101-RandInit & 
0.929 ± 0.024 & 0.803 ± 0.036 & 0.414 ± 0.057 & 0.240 ± 0.059 & 0.885 ± 0.022 & 0.654 ± 0.038\\ 

ResNet101-ImageNet & 
0.971 ± 0.004 & 0.899 ± 0.011 & 0.542 ± 0.022 & 0.355 ± 0.035 & 0.960 ± 0.006 & 0.746 ± 0.005\\ 

ResNet101-BiT-S & 
0.978 ± 0.008 & 0.911 ± 0.004 & 0.581 ± 0.021 & 0.388 ± 0.033 & 0.963 ± 0.002 & 0.764 ± 0.009\\ 

ResNet101-BiT-M & 
\textbf{0.983 ± 0.003} & \textbf{0.919 ± 0.002} & \textbf{0.612 ± 0.022} & \textbf{0.409 ± 0.045}  & \textbf{0.968 ± 0.003} & \textbf{0.778 ± 0.011}\\ 
\hline
\end{tabular}
}

\end{table*}

\subsection{External Validation on Public Dataset}
In external validation, our models showed robust performance to distribution shifts (Figure 4). As the external dataset was more granular (only had two classes, normal and sclerosis), we transformed our model prediction into normal vs. others. Since each model was evaluated using five-fold cross-validation, we averaged the five predicted scores of an image to compute the mean score and the standard error. Again, as shown in Fig 4, the best performing model was ResNet101-BiT-M and it achieved an AUC score of 0.994 when tested directly on the external dataset without any forms of fine-tuning, which was 0.185 higher than training from scratch and 0.048 higher than training using regular ResNet with ImageNet-pretrained weights. In general, BiT models demonstrated major improvements in terms of robustness to distribution shifts. 

\begin{figure*}
  \begin{center}

  \includegraphics[width=0.77\textwidth]{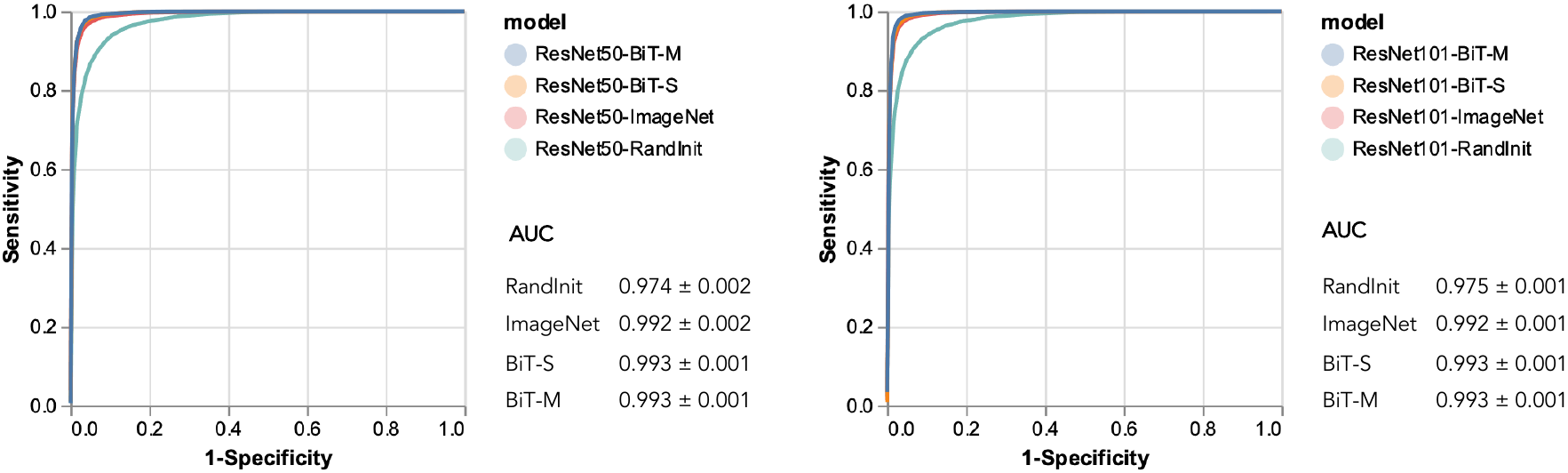}
 
  \end{center}
  \caption  
  {  
     To compare how different models generalize on the external dataset, we provide their ROC curves on the internal dataset. Since our dataset is more fine-grained, we treat sclerosed glomeruli (obsolescent, solidified, disappearing) as the positive class, and non-sclerosed glomeruli (normal, non-glom) as the negative class to perform the analysis. We can see that all these models perform extremely well on the internal dataset to identify sclerosed glomeruli. For both architectures, retrained models outperform models that were trained from scratch. However, the performance difference is minimal between pretrained models.
  }
\end{figure*} 
   
\begin{figure*}
  \begin{center}

  \includegraphics[width=0.8\textwidth]{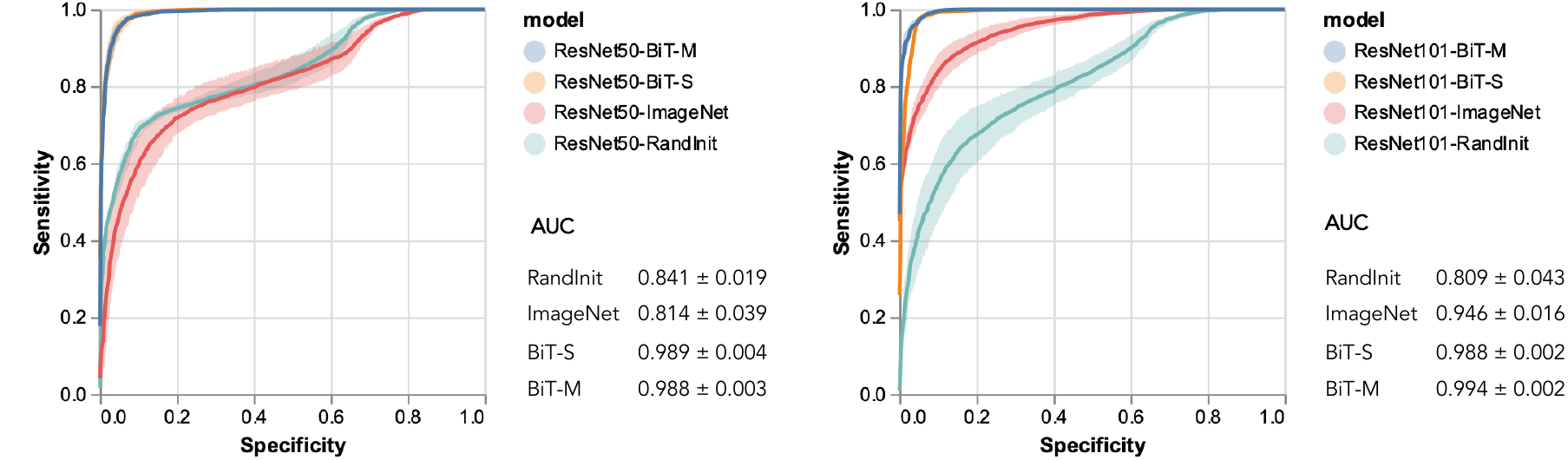}
 
  \end{center}
  \caption  
  {  
     This figure shows the ROC curves (with the AUC scores) of our models on the external validation dataset, whose labels are more coarsely defined (normal vs. sclerosis glomeruli). The error band shows one standard deviation. 
  }
\end{figure*} 

\subsection{Object Detection}

For object detection, our trained classifier can effectively serve as a postprocessing step after any glomeruli detection algorithms to filter out false positives. To demonstrate this application, we show that our classifier consistently improved upon the raw detection results on two CircleNet\cite{yang2020circlenet} models. The first CircleNet model (CircleNet-V1) is selected by training and validating on 704 glomeruli from 42 biopsy samples. Then, this model is used to detect glomeruli on 18 human nephrectomy images, from which 7,449 glomeruli ended up being curated after manual QA by an expert renal pathologist. The second CircleNet model (CircleNet-V2) is then selected by training and validating on all available data. Both models were tested using 1384 manually annotated glomeruli from five unseen human nephrectomy images. The second model performed better in terms of average precision (AP) scores (Table 2 and 3) as it not only learned from more data but also benefited from implicit expert feedback as false positives were removed during the QA. As shown in the bar chart in Figure 5, all of our classifiers had decent AUC scores in identifying non-glomerulus patches and ResNet101-BiT-M performed the best. Therefore, we used a trained ResNet101-BiT-M classifier to filter detection results from CircleNet. The precision-recall plot in Figure 5 demonstrated that this two-stage detection-then-classification pipeline was able to outperform the end-to-end detection solution. We provided detailed average precision results in Table 2 and Table3. The columns denote AP (average precision) , $AP_{50}$ (IOU threshold at 0.5), $AP_{75}$ (IOU threshold at 0.75), $AP_{S}$ (small scale with area $<$ 1000), $AP_{M}$ (median scale with area $>$ 1000). 

\begin{figure*}
  \begin{center}

  \includegraphics[width=0.9\textwidth]{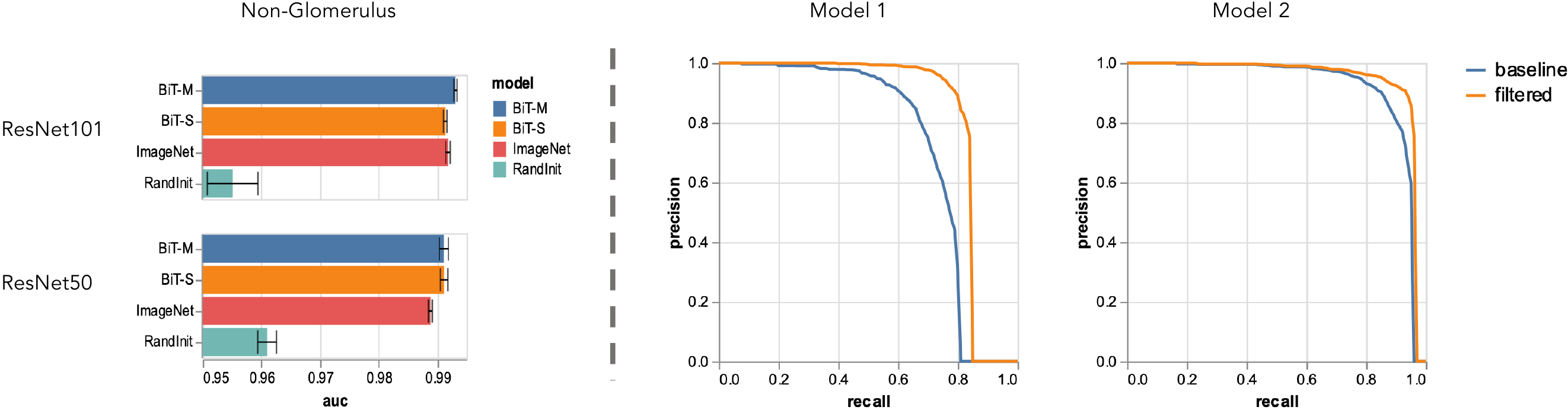}
 
  \end{center}
  \caption  
  {  
     The bar chart on the left panel shows the AUC score for identifying non-glomerulus patches. The right panel shows the precision-recall curves of two glomerular detection algorithms (Model 1 and Model 2) with different performance. In each plot, the blue curve denotes detection results by only using CircleNet, while the orange curve denotes detection results after further using our classifier to filter out false positives. 
  }
\end{figure*}

\begin{table}
    \centering
    \caption{CircleNet-V1 Detection results with/without filtering}
    \medskip
    \begin{tabular}{c c c c c c c}
         \hline
          & Filter & $AP$ & $AP_{50}$ & $AP_{75}$ & $AP_{S}$ & $AP_{M}$\\
         \hline
         CircleNet-V1 & \xmark & 0.504 & 0.729 & 0.511 & 0.363 & 0.721\\
         CircleNet-V1 & \checkmark & \textbf{0.554} & \textbf{0.825} & \textbf{0.550} & \textbf{0.446} & \textbf{0.729} \\
         \hline
    \end{tabular}
    \label{CircleNetOriginalTable1}
\end{table}

\begin{table}
    \centering
    \caption{CircleNet-V2 Detection results with/without filtering}
    \medskip
    \begin{tabular}{c c c c c c c}
         \hline
          & Filter & $AP$ & $AP_{50}$ & $AP_{75}$ & $AP_{S}$ & $AP_{M}$\\
         \hline
         CircleNet-V2 \quad & \xmark \quad & 0.620 & 0.915 & 0.602 & 0.531 & 0.756\\
         CircleNet-V2 \quad & \checkmark \quad & \textbf{0.631} & \textbf{0.941} & \textbf{0.614} & \textbf{0.554} & \textbf{0.757}\\
         
         \hline
    \end{tabular}
    \label{CircleNetOriginalTable2}
\end{table}

\section{Discussion}
In this paper, we presented a deep learning based pipeline to perform holistic GGS detection and classification. To tackle the issue of unbalanced classification, we introduced batch sampler and focal loss to minimize its impact on the optimization process. Transfer learning at scale is demonstrated as a strong paradigm to improve the performance of fine-grained GGS classification. From our experimental results, though they might in fact perform worse than smaller architectures when trained from scratch, larger architectures (ResNet101) turn out to benefit more from transfer learning. For example, in Table 1, while ResNet50-BiT-M performed comparatively to ResNet50-BiT-S, ResNet101-BiT-M showed superiority over ResNet101-BiT-S as well as other models. Thus, larger architectures (ResNet101) are favored over smaller ones (ResNet50) in order to take advantage of transfer learning at a larger scale, as evidenced by both the internal cross-validation results in Table 1 and the external validation results in Figure 4.

Due to the fact that classes such as solidified and disappearing glomerulosclerosis had much fewer samples than the other classes, the classification accuracy of our model on such classes is limited under the multi-class classification scenarios. As there isn't a public fine-grained GGS dataset yet, we employed the dataset that consisted of only normal and GGS classes. Though our model was not particularly optimized for this binary classification task, it was able to achieve superior performance on this external dataset without any forms of fine-tuning, which demonstrated the robustness of our models in the face of distribution shifts. It is worth pointing out that there is an indication of the underspecification phenomenon\cite{d2020underspecification}. In Table 1, all pretrained models seem to be decent when classifying normal glomeruli. In the face of distribution shifts, however, a much larger performance gap is observed, where the BiT models outperformed regular ImageNet-pretrained models by a large margin.

To the best of our knowledge, this is the first work to perform subtypes of global glomerulosclerosis. Previous work has demonstrated deep learning can be deployed to effectively differentiate global glomerulosclerosis from normal using in-house data \cite{uchino2020classification}. Our results strengthen this by demonstrating our algorithm’s performance to identify global glomerulosclerosis from normal on external data. On the other hand, according to Uchino et al.~\cite{uchino2020classification}, the disagreement rate between pathologists in labeling glomerular images can range from 1.7 - 30.9 \%, depending on specific conditions. Since the data we used has been labeled by a single renal pathologist, the study is limited by not capturing the inter-rater variability, which would be valuable for future works. In the future, a following clinical analysis with more than one rater would be critical to understanding the reliability of our trained classifier, which would also help the clinicians to determine the applicability of the proposed method on the clinical application. 

When integrating our classifier with glomeruli detection algorithms, we presented a fully automatic solution to performing quantitative analytics of fine-grained GGS, thus bridging the gap between glomeruli detection and fine-grained GGS classification in the clinical workflow. We hope that the research community can benefit from our open-source solution.

\section{Conclusion}
In this paper, we presented a holistic solution to quantify GGS (both detection and classification) from a whole slide image in a fully automatic manner. In addition, we conducted the fine-grained classification for the sub-types of GGS (obsolescent, solidified, and disappearing glomerulosclerosis). We showed that our proposed holistic solution presented superior results to simple end-to-end object detection algorithms. For fine-grained classification tasks, we demonstrated that transfer learning improved the performance of GGS classification. Moreover, our model has robust performance as it showed superior results when being deployed on an external dataset. In conclusion, we built a highly efficient, open-source, and fully automatic solution for fine-grained GGS Characterization with the raw WSIs as inputs.

\section*{Acknowledgment}
This work was supported by NIH NIDDK DK56942(ABF).

\bibliographystyle{IEEEtran}
\bibliography{main}

\end{document}